\newcommand{\CLASSINPUTbottomtextmargin}{0.75in}
\documentclass[letterpaper, 10pt, conference]{IEEEtran}
\def\IEEEtitletopspace{0.25in}
\IEEEoverridecommandlockouts

\usepackage{cite}

\usepackage{amsmath,amssymb,amsfonts}
\usepackage{algorithmic}
\usepackage{graphicx}
\usepackage{textcomp}
\usepackage{verbatim}
\usepackage{xcolor}
\usepackage{cancel}

\def\BibTeX{{\rm B\kern-.05em{\sc i\kern-.025em b}\kern-.08em
    T\kern-.1667em\lower.7ex\hbox{E}\kern-.125emX}}
\usepackage{ulem}

\usepackage{url}
\usepackage{booktabs}

\usepackage{color}
\definecolor{darkgreen}{rgb}{0,0.5,0}
\long\def\commentml#1{\textcolor{darkgreen}{\bf --MLDM: #1--}}
\long\def\commentml#1{}

\usepackage{eso-pic}

\newcommand{\addITSCheader}{%
  \AddToShipoutPictureFG*{%
    \AtPageUpperLeft{%
      \hspace*{0.7in}%
      \raisebox{-0.4in}{%
        \parbox[t]{\dimexpr\paperwidth-1.50in\relax}{%
          \raggedright
          \sffamily
          \bfseries
          \fontsize{8}{9}\selectfont
          2026 IEEE 29th International Conference on Intelligent\\
          Transportation Systems (ITSC)\\
          September 15--18, 2026. Naples, Italy
        }%
      }%
    }%
  }%
}

\begin{document}

\title{\LARGE \bf Highway Congestion Reduction through \\Reinforcement Learning Based Eulerian Headway Control}

\author{Yaron Veksler$^{1}$, Sharon Hornstein$^{1}$, Han Wang$^{2}$, Maria Laura Delle Monache$^{2}$, and Daniel Urieli$^{1}$

\thanks{$^{1}$Research and Development Labs, General Motors, Herzliya, Israel.
         \newline Email: \{yaron.veksler, sharon.hornstein, daniel.urieli\}@gm.com}%
\thanks{$^{2}$Department of Civil and Environmental Engineering, UC Berkeley. \newline
Email: \{hanw, mldellemonache\}@berkeley.edu
        }%
}
\maketitle

\addITSCheader

\begin{abstract}
%
Connected automated vehicles (CAVs) equipped with adaptive cruise control (ACC) create new opportunities for highway congestion mitigation. Traditional practice relies on \textit{Eulerian} variable speed limits (VSL) which regulate traffic through roadside signs, but suffer from infrequent updates and limited driver compliance.
%
Recent research  explored \textit{Lagrangian} strategies that directly control individual vehicles, offering high reactivity and compliance, yet in realistic multi-lane settings they
depend on
%
drivers’ latent lane-change intentions, 
making robust vehicle-level decisions difficult.
%
Hence,
we propose an \textit{Eulerian} control system optimized through reinforcement learning, that (i) leverages ACC for
reactivity and compliance, and (ii)
obviates dependence on latent driver intentions
%
by regulating aggregate density near bottlenecks,
crucially
via \textit{headway commands} rather than speed commands.
%
%
%
We evaluate two variants of our system, time-headway
and distance-headway control, in large-scale  simulations
across a range of
traffic conditions.
%
Both
variants outperform
baselines,
improving traffic flow by up to 10.6\% over human traffic and 6.7\% over traditional VSL.
%
To strengthen evaluation, we propose a novel boundary-aware speed metric
addressing a recognized flaw in simulation studies with dynamic vehicle entry and exit.
%
%
%
The empirical results, together with our emphasis on deployable system design, suggest a path towards practical, safe, and scalable highway congestion mitigation.
\end{abstract}


\section{Introduction}
\label{sec:introduction}

Highway congestion remains a pressing challenge in modern transportation systems.
The proliferation of connected automated vehicles (CAVs)  equipped with 
adaptive cruise control (ACC)
can be leveraged to
%
improve highway traffic flow and reduce congestion \cite{STERN2018205,10.1109/ICRA.2018.8460567,DelleMonache2019}.
Research on highway congestion mitigation has long distinguished between \textit{Eulerian approaches}, which regulate traffic at fixed roadside locations, and \textit{Lagrangian approaches}, which directly act on individual vehicles~\cite{LAVAL201317}. Common real-world practice relies on traditional Eulerian variable speed limits (VSL) designed for human drivers, where roadside signs display enforced or advisory speed limits.
While studies have demonstrated the potential effectiveness of 
such
systems~\cite{papageorgiou2008effects}, they suffer from two key limitations~\cite{hegyi2005model,han2017variable}: (i) limited responsiveness to dynamic traffic conditions, as roadside signs are sparse and update infrequently (typically only a few times per hour), and (ii) limited and inconsistent compliance by human drivers.

\begin{figure}[t]
    \centering
    \includegraphics[width=\linewidth]{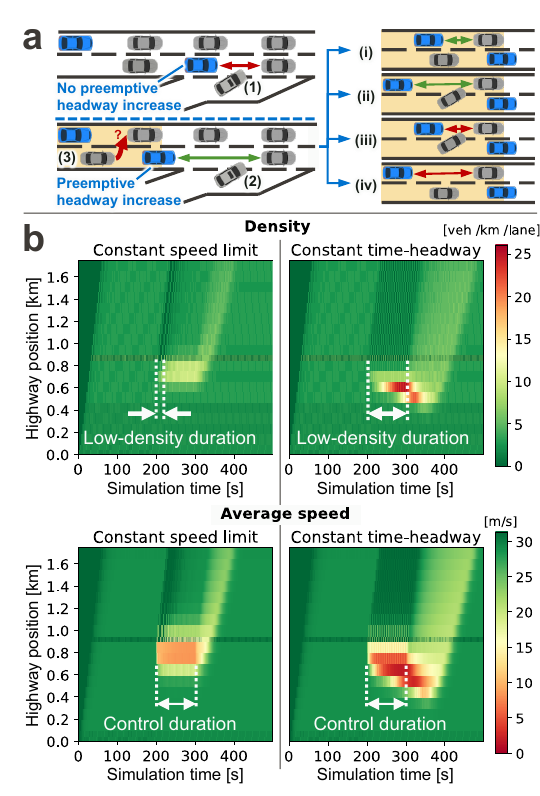}
    \caption{
    \textbf{Headway control motivation:}
    (a)
    Aggressive
    merging
    may cause excessive speed and throughput decrease (1). Preemptively increasing headway can reduce the negative effect (2),
    but may induce lane-changes
    (3). Headway increase decision requires neighboring vehicle
    latent
    intention prediction (i-ii). Wrong prediction causes increased congestion (iii-iv).
    (b) Simulation results for constant speed and time-headway signals. Constant time-headway signals maintain lower downstream density for an arbitrary duration,
    which would otherwise require
    complex dynamic speed limit control (bottom-right).
    }
    \label{fig:qualitative_analysis}
\end{figure}

Advancements in sensing, communication, and automation technologies allow for addressing these limitations. Prior research leveraged these advancements mostly through Lagrangian strategies, which have been
effective in scenarios where the location and time of lane-changes can be accurately predicted, such as in single-lane merge and certain bottleneck scenarios~\cite{Cui2021Scalable,zhang2023learning,vinitsky2020optimizing}. However, Lagrangian strategies struggle with realistic multi-lane highway scenarios~\cite{Bayen2020CIRCLES,lee2024trafficcontrolconnectedautomated},
where lane-change timing and location are highly uncertain and depend on other drivers’ latent intentions and counterfactual responses.
For example, increasing headway can improve flow by enabling cut-ins from congested adjacent lanes, but if those vehicles do not actually merge, the increased spacing merely creates underutilized gaps, and degrades flow (Fig.~\ref{fig:qualitative_analysis}a).
To avoid Lagrangian dependence on latent intentions while maintaining reactivity and compliance, we propose an \textit{Eulerian headway-control system for ACC-equipped vehicles} approaching a traffic bottleneck. Our system continually outputs \textit{headway commands} based on real-time traffic conditions, regulating aggregate density in the bottleneck region. By adopting macro-level density control, our system avoids dependence on unobserved driver lane-change intentions,
and lets vehicles self-organize in sparser traffic.
Moreover, as illustrated in Fig.~\ref{fig:qualitative_analysis}b, regulating density via headway commands is structurally
simpler than the common practice of using speed commands (e.g.~\cite{wang2024hierarchical}).
In our system, reactivity is maintained by continually estimating aggregate traffic information, specifically densities and speeds around bottlenecks, which in the real-world  can be done using modern sensing technologies~\cite{i24motion}.
Compliance is achieved by leveraging ACC-equipped vehicles as actuators,
updating their headways to the values sent by our system.
%
%
 The decision-making component of our system is
 a neural network control policy that maps real-time traffic conditions to
 Eulerian
 headway-control commands, trained with continuous reinforcement learning algorithms such as PPO~\cite{schulman2017proximalpolicyoptimizationalgorithms}.

%
%
%

%
We evaluate two simulated system variants: (i)  time-headway control; and (ii) distance-headway control.
%
Experiments are run
using the  SUMO traffic simulator~\cite{Krajzewicz2012RecentDA}, with thousands of vehicles across a range of merging traffic flows and ACC penetration rates.
%
The two variants
consistently improve traffic flow compared to both human-driven traffic and VSL baselines. Improvements are most pronounced at higher ACC penetration rates, which in the real-world are already substantial and continue to rise each year \cite{parts_adas_2024}, with average speeds up to 
10.6\% higher than human traffic and 6.7\% higher than
traditional VSL.

To improve evaluation quality, we propose a novel average-speed metric for simulated road networks with dynamic vehicle entry and exit, addressing a recognized limitation of existing measures~\cite{Cui2021Scalable}. This metric yields more realistic speed estimates and more reliable comparisons across scenarios.

Our design emphasizes \textbf{practicality}, \textbf{scalability}, and \textbf{safety‑aware operation}: it relies on deployable technologies, aims to operate independently at individual bottlenecks to support scalable deployment, and leverages built-in ACC safeguards. Together, this design and our empirical findings suggest that Eulerian headway control of ACC-equipped vehicles offers a practical, safe, and scalable path to congestion mitigation.
%
Our code, example videos, and supplementary materials are publicly available on GitHub\footnote{\url{https://coopcruise.github.io/}}.

\section{Related Work}
\label{sec:related}

Highway congestion arises from complex driver interactions in dense traffic, motivating extensive research in traffic modeling and control. Foundational work spans macroscopic flow models describing aggregate quantities such as density and flow \cite{Lighthill1955, Hoogendoorn2000}, and microscopic models capturing individual vehicle interactions \cite{Gipps1981, IDM_Treiber}. These models reproduce key traffic phenomena including traffic breakdowns, stop‑and‑go waves, and phantom jams \cite{KERNER2016700, VANWAGENINGENKESSELS2015445}. 
We adopt a microscopic modeling framework that captures the detailed car‑following dynamics, which includes unpredictable drivers intentions or lane‑change behavior.

Among microscopic approaches, the intelligent driver model (IDM)
has been particularly influential due to its ability to reproduce realistic car‑following behavior and congestion patterns \cite{IDM_Treiber, KERNER2016700}.
In our study, we use the SUMO simulator \cite{Krajzewicz2012RecentDA}, whose IDM-based models enable capturing large-scale traffic with heterogeneous driver behavior.

CAVs introduce
opportunities for traffic stabilization. Experiments have shown that even a small fraction of automated vehicles can dampen stop‑and‑go waves and
reduce congestion \cite{STERN2018205}. In mixed‑autonomy
settings,
reinforcement learning (RL) \cite{RL_bookSuttonBarto}
has been
used to derive
CAV
controllers capable of smoothing traffic and improving throughput \cite{10.1109/ICRA.2018.8460567, DelleMonache2019}. RL‑based controllers have been shown to dissipate stop-and-go waves and coordinate merging behavior \cite{Cui2021Scalable}.

A central distinction in CAV control design lies between Eulerian and Lagrangian approaches. Eulerian control strategies typically rely on VSL and have been shown to mitigate congestion under certain conditions \cite{papageorgiou2008effects, hegyi2005model}. Both optimization‑based and data‑driven VSL controllers have demonstrated improvements in throughput and travel time \cite{Li_RL_VSL_2017, Nie_VSL2021}. Nevertheless, their effectiveness is often constrained by low-reactivity, limited 
driver compliance, and by the inherent difficulty of shaping multi‑lane interactions through speed commands alone \cite{doi:10.3141/2423-03}.
%
Lagrangian approaches directly control individual vehicles, maintaining high reactivity and compliance, and have demonstrated strong performance in
scenarios such as single‑lane merges and some bottlenecks \cite{8569485, vinitsky2020optimizing}. The CIRCLES project represented a major step toward real‑world deployment, demonstrating that Lagrangian learning‑based CAV control can be implemented in open‑road experiments at scale \cite{CIRCLES_Link, lee2024trafficcontrolconnectedautomated}. 
However, in realistic multi‑lane settings they are hindered by latent driver intentions and unpredictable lane‑change behavior \cite{DI2021103008},  which make robust vehicle‑level decisions difficult. 
Therefore, traditional Eulerian methods tend to lack responsiveness, while Lagrangian controllers face practical deployment challenges.

These limitations motivate our use of a RL‑based Eulerian strategy that regulates aggregate density rather than relying on vehicle‑level predictions.
%
To address these limitations, we propose a responsive Eulerian framework that regulates traffic density directly through headway control. By modulating headway rather than speed, the controller exerts a more direct
influence on density in the vicinity of bottlenecks and avoids reliance on lane‑change prediction. As demonstrated in our experiments, this headway‑based approach consistently outperforms  both conventional and RL‑based VSL
in complex multi‑lane simulated scenarios. 
\section{Problem Formulation}
\label{sec:problem}

In this section we formalize our congestion-mitigation problem and describe the simulation setting and performance metric used throughout the paper.

\subsection{Congestion-Reduction Objective}

We consider a multi-lane highway
that contains a merging bottleneck. Traffic
consists of
both human-driven vehicles and vehicles equipped with
ACC.
We refer to ACC-equipped vehicles that receive commands from
a
roadside controller as \textit{connected-ACC} vehicles. Let $N$ denote the total number of vehicles
using the road during an evaluation period.
For a given control policy $\pi$, let $\bar{v}_{\pi}^{(i)}$
denote the average speed of vehicle $i$,
and let $\bar{v}_{\text{baseline}}^{(i)}$
denote the corresponding average speed of the same vehicle in a baseline scenario with human-driven traffic and no congestion-mitigation system.
Our goal is to maximize the average relative increase in speed:
\begin{equation}
    J_{\pi} \equiv \frac{1}{N} \sum_{i=1}^{N}
    \frac{\bar{v}_{\pi}^{(i)} - \bar{v}_{\text{baseline}}^{(i)}}
         {\bar{v}_{\text{baseline}}^{(i)}},
    \label{eq:congestion_speed_metric_def}
\end{equation}
subject to traffic dynamics and safety constraints. Since speed reduction is a primary driver of increased travel time, fuel consumption, and emissions, average speed is widely regarded as an appropriate congestion measure~\cite{rao2012measuring}.

\subsection{Traffic Model and Safety Constraints}

We model longitudinal dynamics using IDM for both human-driven vehicles and
connected-ACC
vehicles, consistent with prior work showing that IDM provides a reliable approximation of commercial ACC behavior \cite{10.1109/ITSC.2019.8917347}.
Human-driven vehicles follow IDM with fixed parameters, including a time headway, minimum gap, and maximum speed corresponding to the posted speed limit.
Connected-ACC
vehicles use the same IDM formulation but allow
parameters such as headway or speed set-points
to be adjusted by
an external
controller.

Safety is assumed to be enforced by production ACC systems in the connected-ACC vehicles.
However, since ACC parameters can be modified by the external controller,
its actions are restricted such that, for every vehicle $i$ at all times,
\begin{equation}
    v^{(i)} \leq v_{\max}, \quad
    \Delta^{(i)} \geq \Delta_{\min}, \quad
    \Delta_T^{(i)} \geq \Delta_{T,\min},
    \label{eq:constraints}
\end{equation}
where $v^{(i)}$ is the vehicle speed, $\Delta^{(i)}$ is the
distance headway
to its leader, $\Delta_T^{(i)}$ is its time headway, and $v_{max}$, $\Delta_{min}$, and $\Delta_{T,min}$ are ACC safety speed, distance-headway, and time-headway thresholds.

\subsection{Simulation Platform and Scenario}
\label{sec:sumo}
Our empirical evaluation uses large-scale microscopic simulations of multi-lane highway traffic in SUMO~\cite{Krajzewicz2012RecentDA}. The main highway has four lanes, and a single-lane on-ramp road merges into the rightmost lane, creating a congestion-prone bottleneck with thousands of vehicles engaging in complex car-following and lane-changing interactions. As a
realistic instantiation, we simulate a 2~km segment of I-24 in Tennessee, USA, extracted from OpenStreetMap~\cite{OpenStreetMap}, which has been studied previously as a representative highway-merge geometry~\cite{lee2024trafficcontrolconnectedautomated}.

\begin{figure}[t]
    \centering
    \includegraphics[width=\linewidth]{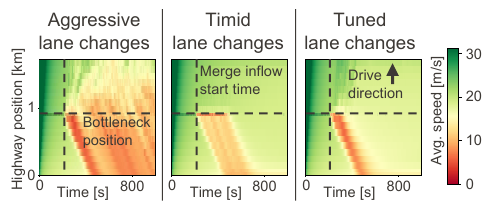}
    \caption{
    Highway dynamics with 25 merging vehicles and maximal inflow under different lane-change behaviors: (left) aggressiveness causes persistent congestion, (middle) timidness causes prolonged right-lane slowdowns, and (right) tuned parameters enable balanced lane use and flow recovery
    }
    \label{fig:lane_change_main}
\end{figure}

Human-driven vehicles are modeled using IDM with empirically tuned parameters that yield a maximum unconstrained inflow of approximately 1800~vehicles/hour/lane.
Lane-changing behavior is modeled with SUMO's
sub-lane model and tuned (e.g., via SUMO's \texttt{lcAssertive} parameter) to avoid unrealistic extremes: excessively aggressive lane changes that create persistent network-wide congestion, and overly timid lane changes that lead to effectively single-lane behavior on the main road. The resulting configuration produces 
plausible
congestion patterns around the bottleneck, as illustrated by time-space diagrams in Fig.~\ref{fig:lane_change_main}.

\subsection{Boundary-Aware Speed Metric for Finite Road Simulations}

In finite-road-length microscopic simulations, where only a bounded highway section is modeled and vehicles enter and exit through the boundaries, naive average-speed metrics can be misleading: a controller may delay vehicle entry, artificially inflating speeds without being penalized for the resulting upstream congestion~\cite{Cui2021Scalable}. To correct this flaw, we define per-vehicle average speeds that explicitly penalize delayed entries.

Let $L^{(i)}$ denote the distance traveled by vehicle $i$ within the simulated road, $T_s^{(i)}$ its planned entry time, $T_f^{(i)}$ its measured exit time, and $T_{\text{sim}}$ the total simulation time. We define
\begin{equation}
    \bar{v}^{(i)} =
    \frac{L^{(i)}}{\min\!\left(T_f^{(i)}, T_{\text{sim}}\right) - T_s^{(i)}},
    \label{eq:average_speed_def}
\end{equation}
so that vehicles that are delayed before entering the network are treated as having speed zero during their waiting time. The baseline and controlled average speeds in~\eqref{eq:congestion_speed_metric_def} are computed using this definition. As a result, strategies that block or delay vehicles cannot achieve high performance by simply keeping the simulated road sparsely populated; they incur a penalty through reduced $\bar{v}^{(i)}$.
Ultimately, the controller design problem can be written as follows.

\vspace{0.2cm}
\noindent\textbf{\underline{Congestion Reduction Optimization Problem}:}
\begin{equation}
\begin{aligned}
    \max_{\pi} \quad J_{\pi}\quad
    \text{s.t.} \quad (v^{(i)}, \Delta^{(i)}, \Delta_T^{(i)}) \text{ satisfy } \eqref{eq:constraints} \quad \forall i,
\end{aligned}
\label{eq:opt_problem}
\end{equation}
with $J_{\pi}$ given by~\eqref{eq:congestion_speed_metric_def} and average speeds computed via~\eqref{eq:average_speed_def}.

\section{Methodology}
\label{sec:methodology}

In this section, we translate the high-level motivation in Section~\ref{sec:introduction} into a concrete methodology. Our approach implements  Eulerian headway controllers operating at the road-segment level, mitigating congestion while satisfying practical requirements of safety, deployability,
and scalability.

\subsection{Control Architecture and Command Types}
\label{sec:control_schemes}

\begin{figure}[t]
    \centering
    \includegraphics[width=0.9\linewidth]{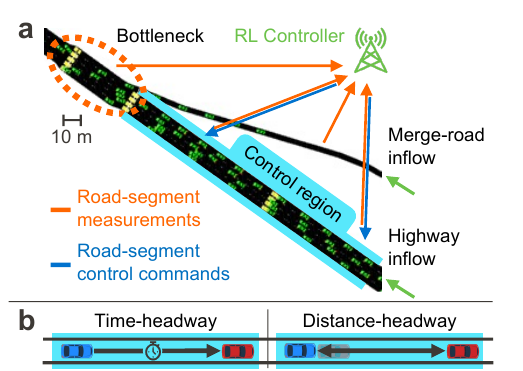}
    \caption{\textbf{Eulerian RL-based headway control for multi-lane highway congestion reduction:} (a) The analyzed scenario. An RL-based controller sends commands to connected-ACC vehicles near bottlenecks, based on measured traffic speed and density. (b) Illustration of examined control schemes:
    time-headway, and distance-headway.
    }
    \label{fig:env_control}
\end{figure}

Our controller continually receives
estimates of average speed and density in each road segment in a region around the bottleneck, and outputs desired  headways, one for each controlled road-segment. Each  headway is broadcast to all connected-ACC vehicles in its corresponding road-segment, and sets their target headway parameter. Hence, decisions are taken in an Eulerian, segment-based manner while actuation is automated
through vehicles' ACC systems
(Fig.~\ref{fig:env_control}a).
We compare
two
control schemes (Fig.~\ref{fig:env_control}b):
\begin{enumerate}
    \item \textbf{Time-Headway Control:}
    Headway is specified as time-gap to leading vehicle.
    \item \textbf{Distance-Headway Control:}
    Headway is specified as distance to leading vehicle.
\end{enumerate}

In all cases, commands are updated frequently, enabling significantly faster adaptation than traditional roadside VSL, which typically adjust only a few times per hour.
\subsection{Reinforcement Learning Formulation and Training}
\label{sec:RL}

To obtain
controllers for each of the control schemes, we
model
the problem as a discrete-time Markov decision process (MDP)~\cite{puterman2014markov} and
approximate its solution
using reinforcement learning.
The solution to an MDP is a control policy mapping states to actions so as to maximize expected cumulative reward; below, we specify the
MDP components
used in our formulation.
\begin{itemize}
    \item \textbf{State/Observation:}
A $2m$-dimensional vector containing
average speeds and densities in a  set of $m$ road segments around the bottleneck (upstream, downstream, on-ramp).

    \item \textbf{Action:} A vector of continuous commands, one per controlled upstream segment, whose semantics depend on
a specific control
scheme, e.g.~time headway or distance headway. To ensure realism and safety, we restrict commands to practical ranges which
are at least as conservative as the default ACC parameters. Control signals are sent to all
vehicles in controlled segments. While lane-specific control could be more effective, it requires complex sensing, limiting practicality.

    \item \textbf{Transition function:} Given by the dynamics modeled by the SUMO simulator.

    \item \textbf{Reward function:}
    Deriving a reward function from the
    objective
    $J_{\pi}$ in~\eqref{eq:congestion_speed_metric_def}
    would lead to highly delayed feedback, since vehicle-level average speeds are only available once routes are completed. Instead, we employ an
    immediate-feedback
    time-delay proxy reward that measures, at each time step $t$, the deviation from free-flow speeds:
\begin{equation}
    r_t = \frac{1}{C}
    \sum_{i=1}^{N_t}
    \left( \frac{v_t^{(i)} - v_{\text{free}}(x_t^{(i)})}
                {v_{\text{free}}(x_t^{(i)})} \right) \Delta t,
    \label{eq:time_delay_reward}
\end{equation}
where $N_t$ is the number of vehicles planned to enter
the simulation by time $t$, $x_t^{(i)}$ and $v_t^{(i)}$ are the position and speed of vehicle $i$, $v_{\text{free}}(x)$ is the local speed limit, $\Delta t$ is the simulation time step ($0.5$s in our case), and $C$ is a
numerical
normalization constant. Vehicles
whose entry to the simulation is delayed due to congestion
are treated as moving at zero speed.
$\sum_{t} r_{t}$
yields a negative quantity proportional to the average travel-time delay relative to free flow, and is a weighted version of $J_{\pi}$.
Therefore, maximizing $\sum_{t} r_{t}$ approximates maximizing $J_{\pi}$.

    \item \textbf{Horizon-length:} The time duration of a full simulation, chosen to allow congestion to develop and dissipate.

    \item \textbf{Decision-interval:} The time between consecutive control signals sent to connected-ACC vehicles, should reflect realistic vehicle response times and acceleration characteristics.
\end{itemize}

\noindent\textbf{RL Training.}
For each control scheme, we train a policy
using
the PPO algorithm
as implemented in RLlib~\cite{pmlr-v48-duan16}, with mostly default hyperparameters. Each training episode corresponds to a single simulation with randomized
selection of the subset of connected-ACC vehicles.
The resulting policies map
the traffic information contained in the current MDP state to the
headway actions
that adapt in real time to evolving congestion patterns around the bottleneck.

\section{Empirical Analysis}
\label{sec:experiments}
\begin{figure*}[t]
    \centering
    \includegraphics[width=0.85\linewidth]{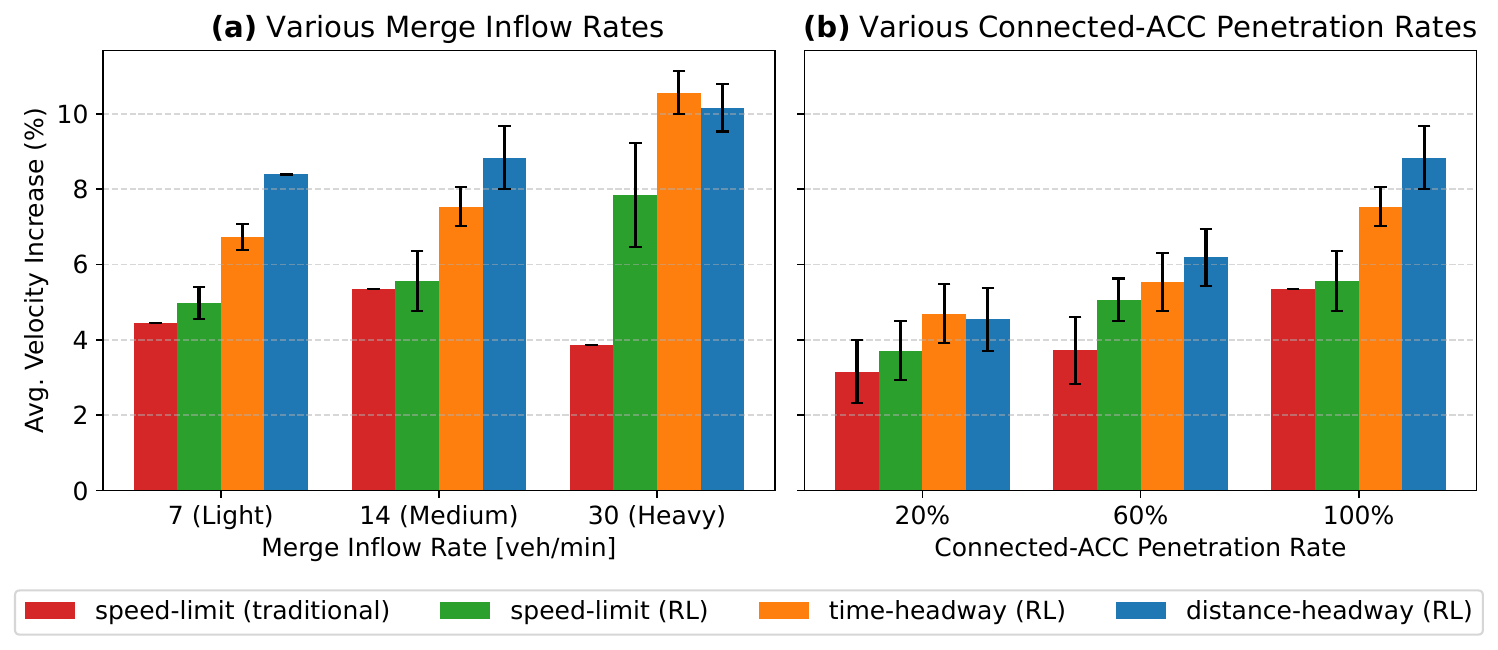}
    \caption{
    \textbf{Time-headway control performance}:
    RL-based Eulerian headway control compared to baseline VSL approaches, under (a) light, medium, and heavy merging traffic with 100\% connected-ACC penetration rate; (b) medium merging traffic with 20\%, 60\%, and 100\% connected-ACC penetration rates.
    }
    \label{fig:results_two_panels}
\end{figure*}
This section describes the experimental setup and empirical analysis of our system, comparing its two variants
with three realistic baselines.
All experiments were conducted in the
simulation environment
described in Section~\ref{sec:sumo}.
Simulations ran for 500 seconds with roughly 1,000 vehicles.
Highway traffic inflow matches
maximum capacity of 1800 vehicles/hr/lane.
Merging vehicles enter after a 60-second warm-up, allowing vehicles to fill the main road and reach steady flow. 
To evaluate performance under varying road conditions, we test three rates of merging inflow: light (7 vehicles/min), medium (14 vehicles/min), and heavy (30 vehicles/min),
and three ACC penetration rates: 20\%, 60\%, and 100\%.
Merging traffic continues for 50 seconds, creating realistic traffic disturbances that impact flow. This design isolates the controller’s response to a localized transient disturbance, enabling us to clearly evaluate both the onset and dissipation of congestion
and assess the controller’s ability to stabilize traffic.
Default ACC parameters are: time-headway of 1.5 seconds, distance-headway of 2.5 meters, and speed limit of 31.29 m/s (70 mph), matching the highway limit.

In all
tested control schemes,  the
state includes 11 main road
segments upstream of (and including) the bottleneck, one on-ramp segment, and 7 main road downstream segments. Each segment spans approximately~100~m, balancing sensing resolution and controller stability.
Actions are two-dimensional continuous vectors, containing commands for two controlled segments immediately upstream from the bottleneck; their semantics depend on
the control scheme, as specified below.
The
decision-interval
was set to 2.5~seconds.  For each scenario, we tested the performance of the following control schemes:

\begin{enumerate}
    \item \textbf{(Baseline) Human-Driven Traffic:} all vehicles are human-driven and follow default behavior with nominal speed limits with no congestion mitigation systems.
    \item \textbf{(Baseline) Traditional VSL:}
    action commands are constant reduced speed limits, activated whenever merging vehicles are detected on the on-ramp segment.
    Human-driven vehicles share the road with connected-ACC vehicles that obey these actions. To approximate the best performance achievable by roadside VSL,
    this constant limit is selected via a grid search, in 10\% increments of the nominal speed limit.
    \item \textbf{(Baseline) RL-based VSL:} connected-ACC vehicles follow RL-based speed-limit action commands 
    in the range $[0,31.29]$ m/s,
    enabling more responsive control policies than traditional VSL.
    \item \textbf{(Ours) RL-based Eulerian time-headway control:} connected-ACC vehicles follow RL-based time-headway action commands in the range $[1.5,6]$ seconds.
    \item \textbf{(Ours) RL-based Eulerian distance-headway control:} connected-ACC vehicles follow RL-based distance-headway action commands in the range $[0,30]$ meters.
\end{enumerate}

Fig.~\ref{fig:results_two_panels}
shows simulation results from thousands of experiments across varying merging inflows and connected-ACC penetration rates,
comparing the performance of Traditional VSL, RL-based VSL, RL-based time‑headway, and RL-based distance‑headway.
Each scenario configuration was run  with 30 different random seeds, from which  95\% confidence intervals were computed.
Fig.~\ref{fig:results_two_panels}a focuses on the case of 100\% connected‑ACC vehicles under light, medium, and heavy merging inflows. Fig.~\ref{fig:results_two_panels}b examines a complementary dimension by fixing the merging inflow at a medium level and varying the connected-ACC penetration rate.
Across tested scenarios, the headway-based controllers outperform baselines; at 100\% connected-ACC penetration rate they increase average speed by up to 10.6\% relative to human-driven traffic, up to 6.7\% relative to traditional VSL, and up to 3.4\% relative to  RL-based VSL.
When fixing the merge inflow, the performance of  headway-based controllers improves as ACC-penetration increases, which is a desirable practical property.
The superior performance of headway-based methods supports the intuition that direct headway-based density control can improve traffic flow in complex multi-lane scenarios, underscoring its potential for scalable, resilient congestion mitigation in connected-vehicle traffic management.

\section{Conclusions}

This paper presented an RL-based Eulerian   system for congestion mitigation at highway bottlenecks that issues headway control commands to ACC-equipped vehicles. The design emphasizes practical deployment, leveraging existing perception, communication, and control capabilities, as well as built-in ACC safety mechanisms. We conducted a large-scale empirical evaluation in multi-lane
simulations across a range of merging flows and ACC penetration rates, showing that headway-based control consistently improves traffic flow relative to human-driven traffic, traditional roadside sign-based speed limits, and RL-based speed-limit control, with the greatest gains observed at higher ACC penetration rates.
Collectively, these results highlight the potential effectiveness, safety, and scalability of  Eulerian headway control for mitigating congestion at multi-lane bottlenecks and improving highway traffic flow.
While these results are encouraging, several avenues for future work remain.
First, deploying an RL controller trained in simulation into the real world requires it to be
trained on simulations calibrated with real-world data. Additionally, real-world testing is crucial to validate the simulation results and overcome practical implementation challenges.
Finally, integrating this approach with other traffic management strategies, such as ramp metering or dynamic lane assignment, could potentially yield even greater congestion reduction.
As automated-vehicle technologies continue to mature, our findings indicate that Eulerian headway control of ACC-equipped vehicles, enabled by existing sensing, communication, and built-in safety mechanisms, offers a practical, safe, and scalable pathway for congestion mitigation.

\bibliographystyle{unsrt}
\bibliography{main.bib}

\end{document}